\documentclass[prl,twocolumn,amsmath,amssymb,superscriptaddress,
showpacs]{revtex4}
\usepackage{graphics}
\usepackage{epsfig}
\usepackage{bbm}
\usepackage[latin1]{inputenc} 

\newcommand{\be}{\begin{equation}}
\newcommand{\ee}{\end{equation}}
\newcommand{\eea}{\end{eqnarray}}
\newcommand{\bea}{\begin{eqnarray}}

\newcommand{\eins}{\openone}

\newcommand{\PP}{\ensuremath{\mathcal{P}}}

\newcommand{\ketbra}[1]{\ensuremath{| #1 \rangle \langle #1 |}}

\newcommand{\ket}[1]{\ensuremath{|#1\rangle}}

\newcommand{\braket}[2]{\ensuremath{\langle #1|#2\rangle}}

\newcommand{\kommentar}[1]{}

\renewcommand{\vr}{\ensuremath{\varrho}}

\begin{document}
\title{Multiparticle entanglement under the influence 
of decoherence}
\date{\today}
\begin{abstract}
We present a method to determine the decay of multiparticle quantum 
correlations as quantified by the geometric measure of entanglement 
under the influence of decoherence. With this, we compare the robustness 
of entanglement in GHZ-, {cluster-,} W- and Dicke states of four qubits and 
show that the Dicke state is most robust. Finally, we determine the 
geometric measure analytically for decaying GHZ and cluster states
of an arbitrary number of qubits.

\end{abstract}

\author{O. G\"{u}hne}
\affiliation{Institut f\"ur Quantenoptik und Quanteninformation,
~\"Osterreichische Akademie der Wissenschaften,
6020 Innsbruck, Austria,}
\affiliation{Institut f\"ur Theoretische Physik,
Universit\"at Innsbruck, Technikerstra{\ss}e 25,
6020 Innsbruck, Austria}

\author{F. Bodoky}
\affiliation{Kavli Institute of Nanoscience, Delft University of Technology, Lorentzweg 1, 2628 CJ Delft, The Netherlands}

\author{M. Blaauboer}
\affiliation{Kavli Institute of Nanoscience, Delft University of Technology, Lorentzweg 1, 2628 CJ Delft, The Netherlands}

\pacs{03.67.Mn, 03.65.Yz, 03.67.Pp}

\maketitle

{\it Introduction ---} The decoherence of quantum states is a
process in quantum dynamics, which is relevant for
the discussion of fundamental issues like the transition
from quantum to classical physics \cite{zurek}. Also from 
a practical point of view decoherence phenomena have to be 
studied, as they occur in experiments involving entanglement 
and their suppression is of vital importance for any 
implementation of quantum information processing.

Due to this importance, the influence of decoherence on 
the entanglement of multiparticle systems has been 
studied from several perspectives \cite{kempe, acin}. These
investigations concerned either the lifetime of entanglement 
or the entanglement properties of the bipartite system which 
arise if the multiparticle system is split into two parts. 
The lifetime of entanglement, however, gives no quantitative
information about the decay of entanglement \cite{acin}. Moreover, 
as a highly entangled multiparticle state may be separable 
with respect to each bipartition \cite{toth}, considering 
bipartite aspects only may not lead to a full understanding 
of the decoherence process. It is therefore highly desirable 
to investigate a full multipartite entanglement measure under 
the influence of decoherence. Unfortunately, all known entanglement 
measures for multiparticle entanglement are defined via complicated 
optimization procedures \cite{plenio}, which makes it practically 
impossible to compute them for a given mixed quantum state.

In this paper  we present a method to investigate the decay of quantum 
correlations which can be used to overcome these difficulties. We study 
different four-qubit states and use our method to compare 
their robustness against decoherence, using a phenomenological model 
described below. Our approach allows to compute the entanglement for
GHZ and cluster states of an {arbitrary number} of qubits and thereby
to investigate the scaling behavior for these states under decoherence.
As we will further see, our results can be directly tested in nowadays
experiments with photons or trapped ions. Finally, from the viewpoint 
of pure quantum information theory, our results represent one of the few 
cases where the computation of a relevant entanglement measure for mixed 
states can be performed \cite{wootters}.

{\it The Situation ---} We consider the following situation:
a pure quantum state $\ket{\psi}$ is
prepared at time $t=0,$ and in the presence of noise evolves
to a mixed state $\vr(t).$ Our task is to quantitatively investigate
the time evolution of the entanglement
$E(t)= E[\vr(t)]$, and its dependence on the initial state
and the number of qubits.

As entanglement quantifier we use the {\it geometric measure of
entanglement} \cite{geo}. This is a popular entanglement
monotone for multiparticle systems, which is related
to the discrimination of multiparticle states with local
means \cite{hayashi} and has been investigated from several 
perspectives \cite{hayashi2, akimasa, orus}.
For pure states, it is defined as
\be
E_G(\ket{\psi}) =
1- \max_{\ket{\phi} = \ket{a}\ket{b}\ket{c}...}
|\braket{\phi}{\psi}|^2,
\label{geo1}
\ee
i.e. as one minus the maximal overlap of $\ket{\psi}$ with fully separable
states $\ket{\phi}$. It is extended to mixed states by the convex roof construction
\be
E_G(\vr) = \min_{p_k, \ket{\phi_k}} \sum_k p_k E_G(\ket{\phi_k}),
\label{geo2}
\ee
where the minimization is taken over all convex
decompositions of $\vr,$ i.e. over all probabilities
$p_k$ and states $\ket{\phi_k}$ which fulfill
$\sum_k p_k  \ketbra{\phi_k} = \vr.$ Clearly, the
optimization in Eq.~(\ref{geo1}) and especially
in Eq.~(\ref{geo2}) is difficult to perform.

Our method can be summarized as
follows: since any set of  probabilities $p_k$ and
states $\ket{\phi_k}$ in Eq.~(\ref{geo2}) results
in a valid upper bound, we obtain a good upper bound
by choosing them appropriately. Then we 
use the
results of Refs.~\cite{grw, grw2} to obtain a lower bound on
$E_G(t).$ There it has been shown how the
geometric measure can be estimated if the mean value
of a single or a few observables is given \cite{nutshell}.
We show that the lower and the upper bound coincide for 
the multiparticle states we investigate below, allowing for a
precise determination of $E_G(t).$

The noise we consider is described by a master equation
for the matrix elements $\vr_{kl}$ as they are used in phenomenological models
of decoherence for e.g.~electron spin qubits \cite{decoherence}:
\be
\partial_t \vr_{kl} =
\left\{
\begin{array}{l l}
\sum_{i \neq k} (W_{ik}\vr_{ii} - W_{ki}\vr_{kk}) &\mbox{  for $ k =l$},
\\
-V_{kl}\vr_{kl} &\mbox{  for $ k \neq l$}.
\end{array}
\right.
\label{mg1}
\ee
We consider a global dephasing process, where the relaxation of the 
diagonal elements plays no role ($W_{ij} =0$ $\forall i,j$) and the 
off-diagonal terms are affected by a global dephasing rate 
$V_{kl}\equiv\gamma.$ This process leads to exponentially decaying 
off-diagonal components $\vr_{kl}(t) = x \vr_{kl}(0),$ where here 
and in the following $x \equiv e^{-\gamma t}.$ At the end of the paper
we will discuss extensions of our method to other (e.g.~local) 
decoherence models.

{\it Four-qubit states ---} Before presenting our results
for different four-qubit states in detail, note 
that by construction the geometric measure is a 
convex quantity, i.e.~$E_G[\lambda \vr_1 + (1-\lambda)\vr_2]
\leq
\lambda E_G(\vr_1) + (1-\lambda) E_G(\vr_2).$
From this, it follows directly that $E_G(t)$ 
is monotonically decreasing. Moreover $E_G(x)$ 
is convex in the parameter $x$ [since $\vr(x)$ 
depends linearly on $x$], and consequently $E_G(t)$ 
is convex in the time $t$.

Let us start our discussion with the four-qubit
Greenberger-Horne-Zeilinger (GHZ) state \cite{ghz}, 
given by
\be
\ket{GHZ_4}=(\ket{0000}+\ket{1111})/{\sqrt{2}}.
\label{ghzdef}
\ee
The geometric measure of the GHZ states equals $1/2$ as the maximal 
overlap with product states is $1/2$ \cite{geo}.

In order to estimate the entanglement from below, note that the
time dependent fidelity is given by
$
F(t) = Tr[\vr(t)\ketbra{GHZ_4}]= (1+e^{-\gamma t})/2.
$
Using the results of Ref.~\cite{grw} 
we can estimate the geometric measure 
from $F(t)$. Explicitly, it has been shown that if for 
an arbitrary $\vr$ the fidelity $F$ of a 
state $\ket{\psi_{0}}$ is given, 
then $E_G$ is bounded by
\be
E_G(\vr) \geq  \sup_{r \geq 0}
\{
1 + r F(t) -\tfrac{1}{2}
[1+r+\sqrt{(1+r)^2-4 r E_G(\psi_{0})}]
\}.
\label{lowerbound}
\ee
Applying this 
to $\vr(t)$ and the fidelity of the GHZ state leads to (using $x=e^{-\gamma t}$)
\be
E_G(\vr) \geq \tfrac{1}{2} (1-\sqrt{1-x^2}).
\label{lbghz4}
\ee

\begin{figure}[t]
\centerline{\includegraphics[width=0.85\columnwidth]{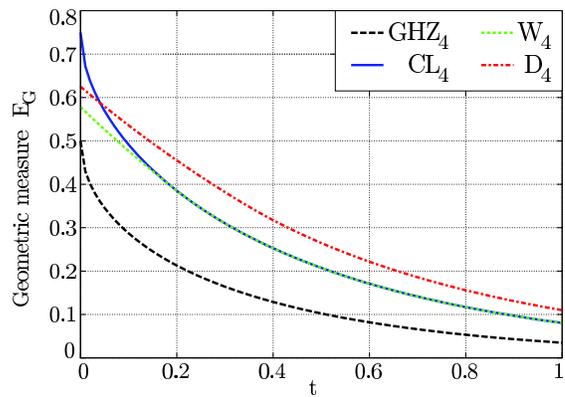}}
\caption{The geometric measure $E_G(t)$ for four-qubit GHZ-, 
cluster-, W- and Dicke states for the case $\gamma=1.$ For 
the W- and Dicke state, the upper bounds are shown, as the curves 
of the lower bounds coincide with that. For $t \gtrsim 2$ the 
values for the W- and cluster state coincide 
[see Eqs.~(\ref{lbcl4}, \ref{geow})].
\label{avbild}
}
\end{figure}

In order to obtain an upper bound, we consider the
two states $\ket{\phi_1} =  c \ket{0000} + s \ket{1111}$
and $\ket{\phi_2} =  s \ket{0000} + c \ket{1111}$ with
$c=\cos(\alpha), s=\sin(\alpha)$, and write
$\vr(t) = (1/2) \sum_{k=1,2} \ketbra{\phi_k}.$ Then,
$x/2 = (2cs)/2$ has to hold 
and using the fact that
the geometric measure for the states $\ket{\phi_k}$
is given by $E_G = \min\{s^2,c^2\}$ \cite{upperboundremark}
one arrives at an upper bound for $E_G(t)$ which is given by the right
hand side of Eq.~(\ref{lbghz4}). Hence,
$E_G(\vr) = (1-\sqrt{1-x^2})/2$ for the four-qubit
GHZ state (\ref{ghzdef}) under the influence
of noise. This function is shown in Fig.~\ref{avbild}.

Second, let us discuss the four-qubit cluster state \cite{clusterstate},
\be
\ket{CL_4}=(\ket{0000}+\ket{0011}+\ket{1100} - \ket{1111})/2,
\label{CL4}
\ee
which has a geometric measure of 
$E_G(\ket{CL_4})=3/4$ \cite{akimasa}.
For this state, the decay of the fidelity is given by
$
F(t)=(1+3x)/4
$,
which [combined with Eq.~(\ref{lowerbound})] leads to the lower
bound
\be
E(\vr) \geq
\tfrac{3}{8}
\big[
1+x -\sqrt{1+(2-3x)x}
\big].
\label{lbcl4}
\ee
For the upper bound, we consider four trial vectors for
the decomposition. The first is given by
$
\ket{\phi_1} = c \ket{0000} + {s}(\ket{1100}+\ket{0011}- \ket{1111})/{\sqrt{3}}
$
and the other three are obtained from this by permuting
the four terms. We choose $c \geq s$, then any of the four
states has a geometric measure of
$
E_G(\ket{\phi_i}) = s^2.
$
With the ansatz $\vr(t)=({1}/{4})\sum_{k=1}^4 \ketbra{\phi_k}$
we obtain as condition on $s$ and $c$ that
$
{x}/{4}=  [({2 c s}/{\sqrt{3}})+ ({2 s^2}/{3})]/4.
$
From this, $c$ can be determined.
This leads after a short calculation to the insight that the right
hand side of Eq.~(\ref{lbcl4}) also constitutes an upper bound on
the  entanglement, and thus describes exactly the time evolution of
the entanglement. 

Third, a four-qubit W state is  given by \cite{duer}
\be
\ket{W_4}= (\ket{0001}+\ket{0010}+\ket{0100}+\ket{1000})/2.
\label{W4}
\ee
for which the geometric measure equals $37/64$ \cite{geo}.
Let us first derive the upper bound. We take as test
states the state
$
\ket{\phi_1} = c \ket{1000} + {s}(\ket{0100}+\ket{0010}+\ket{0001})/\sqrt{3}
$
and permutations thereof. Using a symmetry argument \cite{hayashi2}, 
their geometric measure is determined to be
$E_G (\ket{\phi_i})=[{5+3c^2(c^4+c^2-3)}]/[{(3-4c^2)^2}]$ for
$c \leq {1}/{\sqrt{2}}$ and $E_G (\ket{\phi_i})= 1-c^2$ for
$c \geq {1}/{\sqrt{2}}$ \cite{upperboundremark}. Then,
one can derive an upper bound as for the cluster state.

It turns out, however, that this upper bound is not 
convex in $x$. Since we know that $E_G$  has to be 
convex, we can take the convex hull (in $x$) of this 
upper bound,
\bea
E_G &\leq&
\frac{37(81 x - 37)}{2816} \mbox{ for } x \geq x_0,
\label{geow}
\\
E_G &\leq& \tfrac{3}{8}\big[1+x -\sqrt{1+(2-3x)x}
\big] \mbox{ for } x \leq x_0,
\nonumber
\eea
with $x_0 = 2183/2667.$ Physically, taking the convex hull 
just means that for short times (when $x \geq x_0$) the optimum in
the convex roof in Eq.~(\ref{geo2}) is
of the form $\vr(x) = p \ketbra{W_4} + (1-p) \vr (x_0),$
with $\vr(x_0)=(1/4)\sum_k \ketbra{\phi_k}.$
Note that for longer times ($x \leq x_0$) the upper
bound~(\ref{geow}) is the same as for the cluster state.

In order to see that this upper bound is optimal, let us derive
a lower bound. Here, we not only take the fidelity of the W
state into account, but we use as a second observable the projector
onto the space with one excitation,
$
\PP_1 = \ketbra{0001}+\ketbra{0010}+\ketbra{0100}+\ketbra{1000}.
$
Using the fact that the fidelity of $\ket{W_{4}}$ is given by
$F(t)= (1+3x)/4$ and
that it is always in the space spanned by
$\PP_1$ (i.e., $Tr[\vr(t) \PP_1] = 1$), we can
use the methods of Ref.~\cite{grw} to obtain a lower
bound from these two expectation values
\cite{legendre}. It turns out that, within
numerical accuracy, the lower bound coincides
with the upper bound, giving strong numerical evidence 
that Eq.~(\ref{geow}) is the exact expression for the 
decay of quantum correlations in the W state~(\ref{W4}).

As a last example of a four-qubit state let us
discuss the symmetric Dicke state~\cite{dicke}
given by
\bea
\ket{D_4}&=&(\ket{0011}+\ket{0101}+
\ket{1001}+
\nonumber
\\
&& 
+\ket{1100}
+\ket{0110}
+\ket{1010}
)/{\sqrt{6}},
\eea
which 
has a geometric measure $E_G=5/8$ \cite{geo}.
To obtain the upper bound, we proceed similarly as for the W state:
We take six states, the first one being
$
\ket{\phi_1}={c}\ket{0011}+ {s}(\ket{0101}+
\ket{1001}
+\ket{1100}
+\ket{0110}
+\ket{1010}
)/{\sqrt{5}}
$
and other ones obtained by permuting the terms. Then we make
the ansatz $\vr(t)=(1/6)\sum_{k=1}^6 \ketbra{\phi_k}.$ This leads
to an upper bound which is not convex in $x$,
and subsequently taking the convex hull leads to
\bea
E_G &\leq& \frac{5(3 x - 1)}{16} \mbox{ for } x \geq \frac{5}{7},
\label{geod}
\\
E_G &\leq& \tfrac{5}{18}\big[1+2x -\sqrt{1+(4-5x)x}
\big] \mbox{ for } x \leq \tfrac{5}{7}.
\nonumber
\eea
The lower bound is found analogously to the W state as well, 
with the
projector $\PP_2$ onto the space with two excitations as 
second observable.
The resulting 
bound coincides again with the upper bound, 
giving strong evidence that Eq.~(\ref{geod}) describes the time 
evolution of the entanglement.

\begin{figure}[t]
\centerline{\includegraphics[width=0.85\columnwidth]{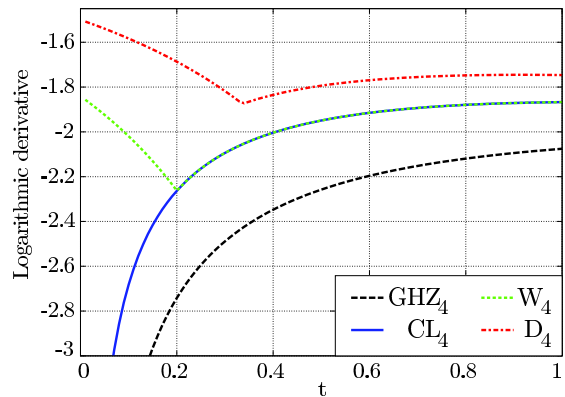}}
\caption{Logarithmic derivative of $E_G(t)$ (for $\gamma =1$) for 
different four-qubit states. The non-analytic behavior for the 
Dicke- and W state stems from the  convex hull 
in Eqs.~(\ref{geow},\ref{geod}).
\label{lnbild}
}
\end{figure}

{\it Comparison of the four-qubit states ---}
For this comparison we
consider the logarithmic derivative
$
\eta = \partial_t (\ln[E_G(t)]) = {\partial_t E_G(t)}/{E_G(t)}
$
which describes the relative decay of entanglement \cite{logder}.
The values of this quantity are plotted in Fig.~\ref{lnbild}.
One can clearly see, that the Dicke state is the most 
robust state, while the GHZ state is the most fragile state 
\cite{fidremark}. 
It is an interesting open question which 
properties of the Dicke state are responsible for the high 
robustness.

{\it Many qubits ---} 
We restrict our attention to GHZ and linear cluster 
states, as they are highly  relevant for applications 
like quantum metrology or measurement-based quantum 
computation \cite{giovanetti}. In the decoherence model, one might 
keep the dephasing rate $\gamma = \gamma_{0}$ constant 
for any number of qubits (where $\gamma_{0}$ is the 
single qubit dephasing rate), or scale it as 
$\gamma = N\gamma_{0}$ (as it would occur for the GHZ state 
in a local dephasing model). For the GHZ state, nothing 
changes 
and all the formulae obtained 
in the above for the four-qubit case apply.
Concerning the cluster state, we consider linear cluster states
with $N=2n$ qubits. The linear cluster state is given by
\be
\ket{CL_{N}}=
\bigotimes_{k=1}^n
{[\ket{00}+\ket{11}(\sigma_x \otimes \eins)]}/{\sqrt{2}},
\label{clustern}
\ee
where this formula should be understood as an iteration, with
the operator $(\sigma_x \otimes \eins)$ acting on the Bell
state of the  next two qubits. Explicitly, we have
$\ket{CL_2}=(\ket{00}+\ket{11})/{\sqrt{2}}$
and
$\ket{CL_4} \sim \ket{00}(\ket{00}+\ket{11}) +
\ket{11} [\sigma_x \otimes \eins (\ket{00}+\ket{11})]
= \ket{0000} + \ket{0011} + \ket{1110} + \ket{1101}$
\cite{clusterremark}.
Note that the maximal overlap of the cluster state with fully
separable states is $1/2^n$ \cite{akimasa}.

\begin{figure}
\centerline{\includegraphics[width=0.85\columnwidth]{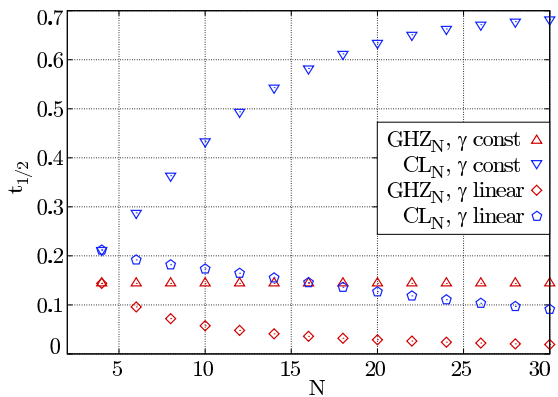}}
\caption{
Comparison between the GHZ state and the cluster
state as a function of the number of qubits $N$. 
The half life time $t_{1/2}$ is shown for the case 
that $\gamma = 4$ and for the case that 
$\gamma = N$ increases linearly with the 
number of qubits.
\label{hwbild}
}
\end{figure}

We calculate the geometric measure for this state 
under the effect of decoherence in the same way as 
for the four-qubit case. The lower bound is obtained 
from the fidelity
$
F(t)=[1+(2^n-1)x]/{2^n}
$
and yields
$
E_G \geq \tfrac{1}{2^N}
\big(
(2-3\cdot 2^n +2^N)x
+ 2 (2^n-1)\{1 - \sqrt{(1-x) [1+(2^n-1)x]}\}
\big).$
For the upper bound, we consider $2^n$ test states
similar to the ones before, and arrive at an upper
bound which coincides again with the lower bound.

To investigate the scaling behavior,
we
consider the time $t_{1/2},$ when the entanglement has decreased
to half of the initial value. These times can be directly computed
for the $N$-particle GHZ and cluster states. Fig.~\ref{hwbild} shows
$t_{1/2}$ as a function of the number of qubits $N$, for the constant and
linear models of $\gamma$. In both cases the time $t_{1/2}$ of the cluster 
state is, in the limit $N \rightarrow \infty$, larger than that of the GHZ 
state by a factor of $\ln(4)/\ln(4/3) \approx 4.82$, 
giving quantitative evidence for the higher robustness against dephasing 
of the cluster state.

{\it Discussion ---} 
In the calculations presented in this paper we concentrated 
on a global decoherence model. However, our results can also be 
applied to other models. First, for the W- and GHZ state, 
our model is equivalent to a local dephasing noise, as it occurs
in multi-photon experiments. Given the experimental availability
of W- and GHZ states with photons \cite{wghz}, our results can be 
directly tested with present-day technology. Second, for local 
decoherence models where relaxation is the dominant process, 
a small modification of our scheme allows the calculation of 
the geometric measure for certain states, such as the W state 
\cite{prep}. The fact that this type of decoherence is dominant
in ion traps \cite{roos} combined with the possibility to generate
such states \cite{haeffner} opens another way for an experimental 
test.

To conclude, our results provide a novel and versatile method 
to determine the decay of multiparticle entanglement for quantum 
states under the influence of decoherence. Our results can be 
tested in multi-qubit experiments and may therefore lead 
to a better understanding of decoherence as a fundamental 
obstacle in quantum information processing.

We thank A. Miyake for discussions. This work has been 
supported by the FWF (START prize), the EU (SCALA, QICS, 
OLAQUI) and by the Netherlands Organisation for Scientific 
Research (NWO).


\begin{thebibliography}{99}
\bibitem{nutshell}
In a nutshell, $E$ can be estimated from the
mean values of two observables $A,B$
as follows~\cite{grw, grw2}: for a given pair
$(\lambda_1, \lambda_2)$ one defines the observable
$X= \lambda_1 A + \lambda_2 B$ and considers its
Legendre transform
$\hat{E}(X) = \sup_{\vr}\{Tr(X\vr)- E(\vr)\}.$
Then, for any state $E(\vr) \geq Tr(X\vr) - \hat{E}(X),$
and more generally
$E(\vr) \geq \sup_{\lambda_1, \lambda_2} Tr(X\vr) - \hat{E}(X).$
The calculation of $\hat{E}(X)$ can often be simplified or
analytically performed \cite{grw, grw2}, and Eq.~(\ref{lowerbound})
is an example thereof.
\bibitem{upperboundremark}
Note that an upper bound of $E_G$ for the trial states
already suffices to derive an upper bound of $E[\vr(t)].$
Such an upper bound is easy to obtain by finding a
product vector with a high overlap in Eq.~(\ref{geo1}).
\bibitem{legendre}
In this case, the calculation of the Legendre
transform can either be done with the algorithm
from Ref.~\cite{grw}, or by direct numerical
optimization, as one can reduce the optimization
to a simple four-parameter optimization
(see Eq.~(25) of Ref. \cite{grw2}).
\bibitem{fidremark}
One may ask at this point whether this conclusion is 
justified, as our decoherence model is not invariant 
under local changes of the basis (as any dephasing model),
leading to different decoherence rates in different
bases. In our case, however, the chosen basis for the
GHZ and cluster state maximizes the long time fidelity
$
F(t = \infty) = Tr[\ketbra{\psi} \vr (t=\infty)] = 
\sum_k \vr_{kk}^2,
$
as the maximal overlap with product states bounds the 
diagonal elements of $\vr$, hence $\vr_{kk}\leq 1/2$ 
(GHZ; resp., $\leq 1/4$ for the cluster state). From 
this it follows that the fidelity in any other basis  
will be smaller at any time, hence lower bounds based 
on the fidelity will only be worse \cite{grw, grw2}. 
This is also a justification 
of the cluster state representation in Eq.~(\ref{clustern}), 
which is in this sense superior to the one in 
Ref.~\cite{clusterstate}.
\bibitem{clusterremark}
One can iteratively check that the resulting state is
an eigenstate of the operators
$S_1 = \sigma_z \sigma_z  \eins \eins ...;
S_2 = \sigma_x \sigma_x \sigma_x \eins \eins...;
S_3 = \eins \sigma_z \sigma_z \sigma_z  \eins ... ;
...;
S_N = \eins ... \eins \sigma_x \sigma_x
$
and thereby fulfills (up to a local rotation) the definition
of a linear cluster state \cite{clusterstate}.
\end{thebibliography}
\end{document}